\documentclass[10pt,preprint2]{aastex61}
\usepackage[parfill]{parskip}    
\usepackage{graphicx}
\usepackage{float}
\usepackage{color}
\usepackage{url}
\pdfoutput=1
\usepackage{threeparttable}

\begin{document}


\title{Physical Conditions in Ultra Fast Outflows in AGN}


\author{S.B. Kraemer}
\affiliation{Institute for Astrophysics and Computational Sciences, Dept. of Physics,
The Catholic University of America, Washington, DC 20064; kraemer@cua.edu}
\author{F.Tombesi}
\affiliation{Department of Astronomy, Univerity of Maryland, College Park, MD
20742,USA; ftombesi@astro.umd.edu}
\affiliation{NASA/Goddard Space Flight Center, Code 662, Greenbelt, MD 20771,
USA}
\affiliation{Department of Physics, University of Rome ``Tor Vergata'', Via della Ricerca Scientifica 1, I-00133 Rome, Italy}
\author{M.C. Bottorff}
\affiliation{Department of Physics, Southwestern University, Georgetown TX,
78626, USA; bottorfm@southwestern.edu}



\begin{abstract}

XMM-{\it Newton} and {\it Suzaku} spectra of AGN have revealed 
highly ionized gas, in the form of absorption lines from H-like
and He-like Fe. Some of these absorbers, ``Ultra Fast Outlows (UFOs)'', have 
radial velocities of up to 0.25c. We have undertaken a detailed photo-ionization 
study of high-ionization Fe absorbers, both UFOs and non-UFOs, in a sample of
AGN observed by XMM-{\it Newton}. We find that the heating and cooling processes
in UFOs are 
Compton-dominated, unlike the non-UFOs. Both types are characterized by Force Multipliers
on the order of unity, which suggests that they cannot be radiatively accelerated in sub-Eddington AGN,
unless they were much less ionized at their point of origin.
 However, such highly ionized gas can be accelerated 
via a Magneto-Hydrodynamic (MHD) wind. We explore this possibility 
by applying a cold MHD flow model to the UFO in
the well-studied Seyfert galaxy, NGC~4151.  We find that the UFO can be accelerated 
along magnetic streamlines anchored in the accretion disk. In the process, we have been
able to contrain the magnetic field strength and the magnetic pressure in the UFO
and have determined that the system is not in magnetic/gravitational equipartition.
Open questions include the variabilty of the UFOs and  the apparent lack of non-UFOs in UFO sources.
\end{abstract}

\keywords{accretion, accretion disks, galaxies: active, X-rays: galaxies}



\section{Introduction}

According to the standard paradigm, Active Galactic Nuclei (AGN) are powered by
accretion of matter onto a supermassive black hole (SMBH). The reservoir of fuel is
thought to be an accretion disk surrounding the black hole, from which 
outflowing winds may arise \citep[e.g.,][]{rees1987}. More than 50\% of Seyfert 1
galaxies, relatively local ($z<0.1$), modest luminosity ($L_{bol} < 10^{45}$
erg s$^{-1}$) AGN,  show intrinsic X-ray and UV absorption \citep{crenshaw2003},
 suggesting that the absorbers have global covering factors $C_{g} \sim
0.5$. Blue-shifted absorption lines in their UV \citep{crenshaw1999}
and X-ray \citep[]{kaastra2000, kaspi2000} spectra reveal significant
outflow velocities (up to $-$4000 km s$^{-1}$, \citealt{dunn2007}). The
inferred mass-loss rates typically exceed the accretion rates needed to produce
the observed luminosities of AGN \citep[e.g.,][]{crenshaw2012}. Hence, mass outflows are a critical
component in the structure, energetics, and evolution of AGN.
Specifically, the relation between bulge mass and black hole mass is thought to be
regulated by AGN outflows, i.e., ``AGN feedback'' \citep{begelman2004}. 
Various acceleration mechanisms have been proposed for these outflows, specifically:
radiative driving \citep[e.g.,][]{murray1995}, thermal winds \citep{begelman1983}, and 
magneto-hydrodynamic (MHD) flows \citep[e.g.,][]{blandford&payne1982, fukumura2010, chakravorty2016}.

If AGN-driven outflows are an important feedback mechanism, they must be
energetic enough to clear the gas in the bulge of the host galaxy and quench
star-formation. Their strength can be expressed in the form of ``Kinetic
Luminosity'', $L_{KE} = \frac{1}{2} \dot{M_{\rm out}}$ $v_{\rm r}^{2}$, where the mass outflow
rate $\dot{M_{\rm out}} = 4 \pi r N_{H} \mu  m_{p} C_{\rm g}  v_{r}$, and $r$ is radial distance of the gas,
$N_{H}$ is column density, $\mu$ is the mean atomic mass per proton ($=1.4$ for solar
abundances), $m_{p}$ is the proton mass, 
$C_{\rm g}$ is the global covering
fraction of the gas, and $v_{r}$ is radial velocity.  
For effective feedback, $L_{KE}$ $\sim$ 0.5\% - 5\% of $L_{\rm bol}$, the
bolometric luminosity of the AGN \citep[]{hopkins&elvis2010, king&pounds2015}.
Since $L_{KE}$ $\propto$ $v_{r}^{3}$, the amount of kinetic energy deposited
into the host galaxy rises quite rapidly with velocity. One caveat is that
the theoretical models for feedback require that the AGNs are radiating close
to their Eddington limit \citep[e.g.,][]{king&pounds2015}, or $L_{\rm bol}/L_{\rm Edd} \sim 1$ (but see below).

\citet{crenshaw2012}~and~\citet{crenshaw2015} have explored the impact of feedback from outflowing
UV and X-ray absorbers and optical emission-line gas, in the Narrow Line Region (NLR),
of a sample of nearby Seyfert galaxies.  For half
of the sample, $L_{KE}$ $\leq 0.5\% L_{\rm bol}$, and most of these sources are significantly sub-Eddington,
hence cannot effectively drive feedback. In their study of the Seyfert 2 galaxy, Mrk 573,
\citet{fischer2017} find that,
while the AGN is radiating at near the Eddington limit, gas is not being radiatively accelerated
at radial distances $\gtrsim$ 1 kpc. These results call into question the effectiveness of AGN
feedback, at least in the local Universe. 
 
However, there are more extreme phenomena, so-called ultra-fast outflows (UFO), which are 
defined as massive, highly ionized, modestly relativistic outflows. They are
identified by narrow Fe K-shell blueshifted absorption lines from Fe XXV/XXVI, 
with $v_{\rm r} \sim$ 0.03 - 0.3c (\citealt{chartas2003,
pounds2003,reeves2003,tombesi2010}). The lines are quite prominent, with EWs in the range 10 - 100 eV 
\citep{tombesi2010}.
UFOs might drive a significant amount of mass and, most importantly, 
energy outwards, and therefore could be a critical component of AGN
feedback.  Photoionization
modeling of UFOs predicts column densities on the
order of N$_{\rm H}$ $\sim$ 10$^{23}$ - 10$^{24}$ cm$^{-2}$, and very 
high ionization, in the range of log $Xi$\footnote{the ionization
parameter $Xi = L_{\rm ion}/n_{H} r^{2}$, where $L_{\rm ion}$ is the ionizing luminosity of the AGN, $n_{H}$ is
the hydrogen number density, and $r$ is radial distance \citep{tarter1969}. We use
$Xi$ rather than the Greek letter $\xi$ to avoid confusion with the scaling parameter used for 
self-similar MHD solutions.} $\sim$ 3 - 6.
There are also highly ionized absorption components that do not fit within the 
UFO parameterization, showing log ($Xi$) $<$ 3 and 
$v_{\rm r} <$ 0.03c, which are classified as non-UFOs
\citep{tombesi2010}.

Previous studies \citep[]{cappi2006,tombesi2010,tombesi2011,tombesi2012,tombesi2014} have 
identified UFOs in archival {\it XMM-Newton} observations of samples of both 
radio-quiet and radio-loud galaxies. Many of them were later confirmed in the
same sources
in {\it Suzaku} spectra by 
\citet[]{gofford2013,gofford2015}. These authors 
have derived qualitative information on the UFO spectral characteristics, 
kinematics and possible location. \citet{tombesi2012} noted that there
seem to be tight correlations between the location of UFO with respect to the 
SMBH, and the ionization state, column density and velocity of the outflowing
gas. The high state of ionization combined with relativistic velocities suggests
an origin in the inner accretion disk. Furthermore, \citet{king&pounds2015} argue that UFOs 
are required for AGN feedback. For example, based on 3-dimensional hydrodynamic simulations, 
\citet{wagner2013} 
suggested that UFOs lose their dependence on opening angle upon their initial
interaction with the interstellar medium of the host galaxy, hence produce
larger-scale feedback than lower-velocity winds (also, see \cite{asahina2017}).

Despite intense study, the origin and 
acceleration mechanism of UFOs are still unclear. There is evidence for variability on timescales
of years \citep[]{reeves2008,pounds&reeves2009,cappi2009}, or as short as days, as in 
the case of Mrk 766 \citep{tombesi2010}, which suggests an origin close to the AGN (although it is posssible
that the variability is due to instabilities within the absorbers
\citep[e.g.,][]{takeuchi2013}).
\citet{gofford2013} and
\citet{king&pounds2015}
argue for acceleration by radiation. However, given their high ionization state \citep{tombesi2011,
gofford2013}, 
this can only occur via electron scattering, which is not likely as most AGN with UFO detections
are radiating at a small fraction of Eddington, unless there are  multiple scatterings, which are not feasible 
given the UFO column densities \citep[e.g.,][]{tombesi2011}. 

On the other hand, it has been suggested that radiative acceleration of UFOs
via UV line-driving is possible if they were in a sufficiently low ionization
state near their launch points
\citep[e.g.,][]{hagino2015, hagino2017, nomura2016, nomura2017}. In this scenario,
as the UFOs flow outwards, they become increasingly ionized, until only H- and
He-like Fe lines can be detected. These models predict that
fast flows can be generated in sub-Eddington sources.

However, it is also plausible that there is a non-radiative means of acceleration, specifically through a magnetohydrodynamic 
(MHD) wind \citep[e.g.,][]{fukumura2010, fukumura2014, chakravorty2016}.
In this paper, we explore this possibility as follows. First, we perform a photo-ionization modeling analysis of the UFOs, and non-UFOs, studied by
\citet{tombesi2010}. In doing so, we obtain constraints on physical conditions within the absorbsers, including:
density, electron temperature, and the heating and cooling processes at work. We then
review the cold MHD flow model
proposed by \citet[][hereafter BP82]{blandford&payne1982}. Finally, in the case of NGC~4151,
a Seyfert galaxy for which we have tight constraints on black hole mass, luminosity, and inclination, we
show that the UFO could be part of an MHD-driven outflow.

\begin{figure}[htb]
\hspace{-1.0cm}\includegraphics[scale=0.35,angle=180]{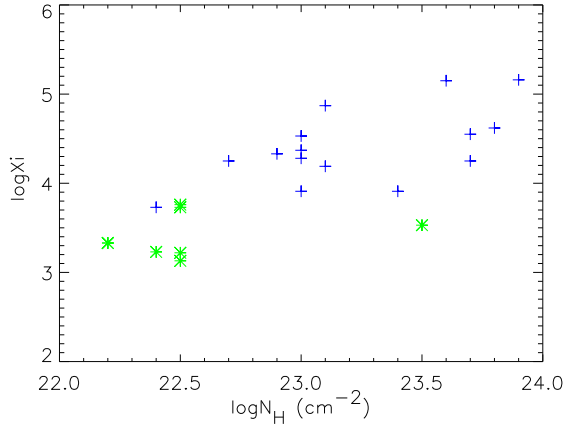}
\caption{\label{fig:f1} Log Xi versus N$_{H}$, for the photo-ionization models
discussed in Section 2; blue crosses, UFOs; green asterisks, non-UFOs. 
These results show that the UFOs and non-UFO occupy different regions 
of parameter space.}
\end{figure}

\begin{figure}[htb]
\hspace{-1.0cm}\includegraphics[scale=0.35,angle=180]{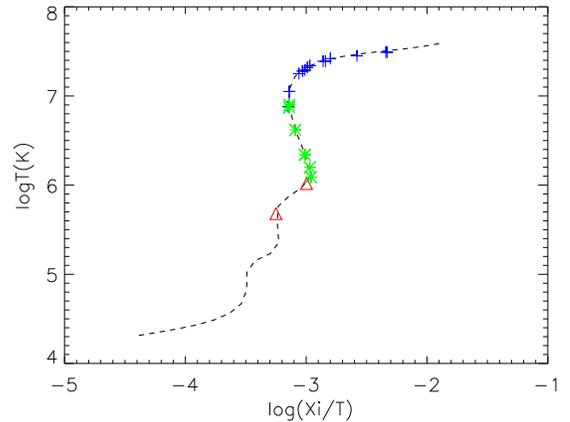}
\caption{\label{fig:f2}Thermal Stability( S-Curve) for the photo-ionization
 models; same as above with red triangles showing two states of the highest 
 ionization warm absorber in NGC 4151 (``XHIGH''; \citet{couto2016}). Note 
 that the three types of absorbers occupy different regions of the S-curve, 
 with the UFOs primarily on the flat, Compton-dominated section.}
\end{figure}

\section{Photionization Models}
\subsection{Model Inputs}

In order to characterize the absorbers, \citet{tombesi2011}~and~\citet{gofford2015} fit spectra
using photoionization models generated with XSTAR \citep{kallman2004}. From the models,
they were able to constrain the range in ionization parameter and column density of the individual
absorbers. As noted above, the models were parameterized in terms of $Xi$.
 
Our priniciple goal is to derive the physical conditions
within the absorbers. To do so, we generated photo-ionization models using Cloudy \citep{ferland2013}, from which
we were able to constrain quantities such as electron temperature, $T_{\rm e}$, relative
contributions to heating and cooling, including Compton processes, and the ``Force Multiplier (FM)'', the ratio of the total 
photo-absorption cross-section to the Thomson cross-section, in a physically self consistent manner. Models
were optimized to match the derived Fe~XXV and Fe~XXVI column densities. These were determined from the
the line Equivalent Width values reported in Table A.2 of \citet{tombesi2010} using the curve of growth analysis described
in \citet{tombesi2011}. We used only lines with an a unambiguous identification as Fe~XXV or Fe~XXVI. We assumed line broadening
due to a turbulent velocity of 1000 km s$^{-1}$ and 3000 km s$^{-1}$ for those associated with  non-UFOs and UFOs, respectively. This
is consistent with the upper limits obtained from the spectral fits in their Table 3. However, 
the broadening may, in fact, be due to velocity gradients along the line-of-sight through the absorbers, rather
than micro-turbulence, a point we will revist in Section 3.2.2. Therefore, we did not include
micro-turbulence in the Cloudy models.

The model results depend on the choice of input parameters, specifically:
the spectral shape of the incident radiation or spectral energy distribution (SED), the radial distances of the emission-line gas with respect to the
central source, $n_{\rm H}$,  and column density ($N_{\rm H}$) of the gas, and its chemical composition. 
We assumed an SED similar to that used by \citet{tombesi2011}, in the form of a power law F${_\nu}$ = K ${\nu^{-\alpha}}$, with 
$\alpha$ = 1.0 for 1.4 $\times 10^{-4}$ eV $<$ h$\nu$ $<$ 100 keV, with exponential cutoffs above and below the limits. This
SED is simpler than the broken power-law \citep[e.g.,][]{laor1997} that we have used in our modeling of Seyfert
spectra \citep[see][]{couto2016}, but its use maintains consistency with the previous UFO analysis.
As in our most recent warm absorber studies
\citep[e.g.,][]{couto2016}, we have assumed roughly 1.5 times solar elemental abundances \citep[e.g.,][]{asplund2005}
as follows (in logarithm, relative to H, by number):
He: $-1.00$, C: $-3.47$, N: $-3.92$, O: $-3.17$, Ne: $-3.96$,
Na; $-5.76$, Mg: $-4.48$,  Al: $-5.55$, 
Si: $-4.51$,  P: $-6.59$, S: $-4.82$,  Ar: $-5.60$, 
Ca: $-5.66$,  Fe: $-4.4$, and Ni: $-5.78$. 

Ionic columns densities were fit by adjusting $Xi$ and $N_{\rm H}$ within the constraints
determined by \citet{tombesi2011}. For a given $Xi$, $n_{\rm H}$ is a function of radial
distance, $r$, and the ionizing luminosity, $L_{\rm ion}$; for the latter, we assumed the values
from \citet{tombesi2012}. We set the upper limits for the radial distance by
requiring that the physical depth, $\Delta r$,
not exceed $r$, or $\Delta r/r <$ 1. Noting that $\Delta r$ $=$ $N_{\rm H}$/$n_{\rm H}$, 
from the definition of  $Xi$ we obtain the expression
$r \geq L_{\rm ion}/(Xi~N_{\rm H})$. Then, by substituting this back into the definition of $Xi$,  $n_{\rm H} \leq  Xi N_{\rm H}^2/L_{\rm ion}$.

\subsection{Model Results}

Model parameters are listed in Table 1 and predicted Fe~XXV and Fe~XXVI column
densities, compared with the measured values, are shown in Table 2.\footnote{We were not able to obtain a satisfactory model-fit for objects with
only Fe~XXV-detected UFOs, 1H0419-577, NGC 7582, and PG 1211$+$143, or the second observation of NGC~4051, for which the measured Fe~XXVI column
was inconsistent with the constraint on N$_{\rm H}$ from \citet{tombesi2011}.}The model-predicted ionic column densities
are all within a factor of 2 of the measured values, which we deem sufficiently accurate
given uncertainties in the iron abundance. 
As shown in Table 1, Cloudy models predict that both UFOs and non-UFOs have
large columns, $N_{\rm H} \sim 10^{22-24}$ cm$^{-2}$, of highly ionized (log$Xi
= 3.13 -5.16$) gas.
However, there is a difference between the two classes. Non-UFOs generally possess
smaller $N_{\rm H}$ and $Xi$, as is shown clearly in Figure 1. The one outlier among non-UFOs, in terms
of $N_{\rm H}$, is ESO 323$-$G77. This is due to the fact that the Fe~XXV and Fe~XXVI
columns densities in this object are quite large (see Table 2). For the UFOs, the model prediction for the first observation 
of Mrk~766 places it with the non-UFOs in Figure 1. However, the Fe~XXVI column densities are nearly the same
in both observations, there may be some uncertainty in the range in $N_{\rm H}$ \citep[see][]{tombesi2011}.   

As shown in Table 1, Compton heating is the dominant mechanism for UFOs, with cooling due to
Compton and free-free processes. For non-UFOs, while Compton heating can sometimes dominate, heating via
ionization becomes important, and cooling via emission-lines dominates for the lowest ionization
cases, e.g., NGC~3783. The difference in heating and cooling processes among UFOs and non-UFOs is clearly illustrated
in a thermal stability plot (``S-Curve''; Figure 2), in which the UFOs lie primarily on the flat Compton-dominated
section. The non-UFOs are found along the negatively sloped portion of the
curve, suggesting that they are thermally unstable \citep[e.g.,][]{bottorff2000}. However, the shape of the vertical section of the S-Curve depends
on the SED, which is unusually flat for these models (see above) and the atomic data, which may be
incomplete for M- and L-shell iron ions which dominate cooling in these conditions \citep[e.g.,][]{kraemer2015}.
Also, as suggested by \cite{bottorff2000}, a sufficiently strong magnetic
field could stabilize the gas (see their Appendix A3). At a minimum,
this result illustrates the difference in the physical conditions within the non-UFOs 
as compared to the UFOs. In Figure 2 we have also plotted the position of the highest ionization component of warm absorption 
in NGC~4151, ``XHIGH'' \citep[see][]{couto2016}, modeled for two continuum flux states. As suggested by
\citet{tombesi2011}, there is a continuum between UFOs, non-UFOs, and the highest ionization warm absorbers.

Cloudy model predictions for the FM provide constraints on the acceleration mechanism for the absorbers.
As shown in Table 1, the FMs for the UFOs are all close to unity, while the maximum value for the non-UFOs
is 1.7. This indicates that the main source of opacity is electron scattering and, therefore,
radiative driving will be inefficient, unless the sources are radiating at Eddington
or there are multiple scatterings \citep[see][]{gofford2015}, which will not be possible for the
predicted column densities. However, we cannot rule out the possibility that the UFOs were in a
much lower ionization, hence characterized by much larger FMs, at their launch points, as noted above.

\subsection{Regarding Induced Compton Scattering}

As discussed above, the electron temperature in a UFO, $T_{e}$, is set by the balance of heating, due to Compton
scattering, and cooling, due to inverse Compton scattering, with some contribution from free-free scattering. If only Compton 
processes are involved, the electron, or ``Compton'', temperature is proportional to the average photon energy
\citep[e.g.,][]{krolik1981}. If the incident continuum flux is strong enough that the photon occupation number is high,  
and the source is anisotropic, induced Compton scattering can become important
\citep{kompaneets1956}. Unlike inverse Compton scattering, all
photon-electron interactions in induced Compton scattering result in an increase in the electron's kinetic energy. As a result, the electron temperature
can far exceed the Compton temperature \citep{levich&sunyaev1970}. 

Induced Compton scattering becomes important if we consider the minimum radial distances of the UFOs, $r_{\rm min}$. \citet{tombesi2012} computed $r_{\rm min}$
assuming that UFOs are at distances at which $v_{\rm r}$ equals the escape velocity. Under these conditions, Cloudy
models predicted significant contributions from induced Compton scattering. However, Cloudy calculates the induced Compton heating based on
the formalism in \citet{levich&sunyaev1970}, which is only valid if $kT_{e}$ $<<$ $m_{e}$c$^{2}$, where $k$ is Boltzmann's constant
and $m_{e}$ is the electron mass. The result is that $T_{e}$ will continue to increase with photon occupation number, without
any physical limit. However, as discussed in \citet{sazonov&sunyaev2001}, as electrons become increasingly relativistic,
the efficiency of energy transfer via induced Compton scattering drops and the maximum $T_{e}$ generally will not exceed 10$^{9}$K. 
Since Cloudy only considers the non-relativistic limit, the model predictions are not valid in the induced Compton regime. Therefore,
we were not able to constrain the physical conditions of the UFOs at $r_{\rm min}$.

\section{Magneto-Hydrodynamic Flows}

The physics of magneto-hydrodymanic (MHD) outflows from accretion disks has been
described in detail by BP82, for a `cold' MHD flow, in
which magnetic pressure exceeds gas pressure, primarily in the context of
relativistic jets. For radio-quiet AGN, i.e., Seyfert galaxies, similar MHD models 
have been invoked to explain the dynamics of broad emission line clouds 
\citep[][hereafter, EBS92]{emmering1992}~and~warm/UV~absorbers~\citep{bottorff2000}. 
However, most warm absorbers are characterized by FM
$>>$ 1, hence it is likely that acceleration by radiation
pressure is dominant \citep[e.g.,][]{couto2016}. On the other hand, Cloudy
models predict FMs $\approx$ 1 for all of the UFOs and most of the non-UFOs
analyzed here. As noted above, \citet{gofford2013} suggested the possibility of
radiative acceleration of UFOs via electron-scattering. This would require
multiple scatterings, which would not be expected for the column densities of the
UFOs, which are  $< 10^{24}$ cm$^{-2}$. Therefore, magnetic acceleration is a possible
mechanism. Also, \citet{fukumura2014} have suggested that MHD winds can account for the
properties of both UFOs and warm absorbers, in the form of a radially stratified
wind. 

\subsection{Parameterization of cold MHD flows}

The MHD wind model developed in BP82 consists of a self-similar axisymmetric
flow.  Self-similarity is achieved by parameterizing the cylindrical poloidal 
components of the position vector ${\bf r}$ in such a way that flow lines, 
originating in the Keplarian disk at radial ``footprint'' $r_{\rm o}$, are 
easily traced \citep{bottorff2000}. This is achieved by invoking a 
dimensionless free parameter $\chi$ and a dimensionless function $\xi(\chi)$ 
so that the position vector ${\bf r}$, in cylindrical coordinates $r$, $\phi$, 
and $z$, is given by

\begin{equation}
{\bf r} = [r_{\rm o}\xi(\chi),\phi, r_{\rm o}\chi],   
\end{equation}

 (BP82). By varying $\chi$ the set $(r, z) = (r_{\rm o}\xi(\chi),r_{\rm  o}\chi)$
traces the poloidal portion of flow streamlines (and magnetic field lines)
having the footprint $r_{\rm o}$. $\chi=0$ corresponds to the disk plane 
so it is required that $\xi(0) = 1$.

As shown in BP82, the velocity vector along a flow 
streamline 
 is given by:
\begin{equation}
{\bf v} = [\xi^{'}(\chi)f(\chi),g(\chi),f(\chi)]\Big{(}\frac{GM}{r_{\rm
o}}\Big{)}^{1/2}
\end{equation} 
where the prime denotes differentiation with respect to $\chi$, $M$ is the BH
mass,
and the square-root term is the Keplerian velocity at the disk. 
The functions $f(\chi)$, $g(\chi)$, and $\xi(\chi)$ can be determined iteratively 
from the ``cold'' MHD flow equations (see Section 2 of BP82). The flow starts
in a Keplarian orbit, so $g(0) = 1$, and the initial poloidal velocity is zero, 
so $f(0) = 0$.  

An alternative approach for finding the functions $f(\chi)$ and $g(\chi)$ is to
assign $\xi(\chi)$, as in EBS92, and then $f(\chi)$ and $g(\chi)$ are found semi-analytically 
and self-consistently with the choice of $\xi(\chi)$ and the ``cold'' MHD flow
equations. The form of $\xi(\chi)$ used by EBS92 is:

\begin{equation}
 \xi(\chi) = \big{(}\frac{\chi}{0.5{\rm tan}\theta_{\rm o}}~+1\big{)}^{1/2}.
\end{equation}

where $\theta_{\rm o}$ is the launch angle of the flow with respect to the disk.
This form of $\xi(\chi)$ has a value of 1 at $\chi = 0$ and gives the 
poloidal part of the flowlines (and magnetic field lines) a parabolic shape. 
While this approach limits analysis to one class of solutions possible from 
BP82 it illustrates relevant physical characteristics of MHD outflows 
so is adopted here. 

To fully characterize the flow along a streamline five inputs are required: $M$,
$r_{\rm o}$, $\theta_{\rm o}$, $\lambda$, and $\kappa$. The constant $\lambda$ is the ratio 
of the total, specific angular momentum, in both matter and magnetic field, to 
the specific angular momentum in the disk at $r_{\rm o}$  (Equations 2.2 and
2.7b of BP82). The constant $\kappa$ is given by equation 3.13 of EBS92. 
Once $f(\chi)$ and $g(\chi)$ have been determined, the square of the Alfv\'{e}n
Mach number, $m$,
is determined using equations 3.14 and 3.15 in EBS92, and the density 
along the streamline and the magnetic pressure, $P_{\rm mag}$,
are obtained from equations 3.16 and 3.17 of EBS92, respectively. 

Key to constraining the input parameters are two relationships from \citet{bottorff2000}.
The first is a linear relationship between the foot-print radius 
$r_{\rm o}$  and the magnitude of ${\bf r}$. The conversion from one to 
another is a function of angles $\theta_{\rm o}$ and $i$, the angle between the observer's line-of-sight
and the disk axis. It can be written as:
\begin{equation}
r_{\rm o} = |{\bf r}|~{\rm sin}(i)(\sqrt{1+{\rm cot}^{2}(i){\rm
cot}^{2}(\theta_{\rm o})}-{\rm cot}(\theta_{\rm o}){\rm cot}(i)) 
\end{equation}

The second relationship is the projected line-of-sight velocity of the flow:

\begin{equation}
v_{\rm obs} = \Big{[}\xi^{'}(\chi)f(\chi){\rm sin}i~+~f(\chi){\rm
cos}i\Big{]}\sqrt{\frac{G M}{r_{\rm o}}}.
\end{equation}

If we have constraints on $M$ and $i$ and $|{\bf r}|$, we can obtain the footprint
radius, $r_{\rm o}$ for a given $\theta_{\rm o}$. Then with $M$, $i$, and
$r_{\rm o}$, and $v_{\rm obs}$, we can determine the value of $f(\chi)$ (at 
the location of the UFO). But along the line-of-sight $\chi = z/r_{\rm o}=
|{\bf r}|{\rm cos}(i)/r_{\rm o}$, so $\chi$ is determined. The input parameters $\lambda$ and $\kappa$ can
be adjusted until the model reproduces the observationally inferred $f(\chi)$
at the observationally inferred $\chi$.This fixes the flow solution enabling 
estimation of the spatial extent of the absorption system, given the velocity 
width of observed spectral features and the column density. In the next
subsection, we apply this form of a ``cold'' MHD solution to the UFO in NGC~4151, for which
we use values for $M$ and $i$ from previous studies and the constraint on
$|{\bf r}|$ from our photo-ionization analysis.

\begin{figure}[htb]
\hspace{-1.0cm}\includegraphics[scale=0.35,angle=180]{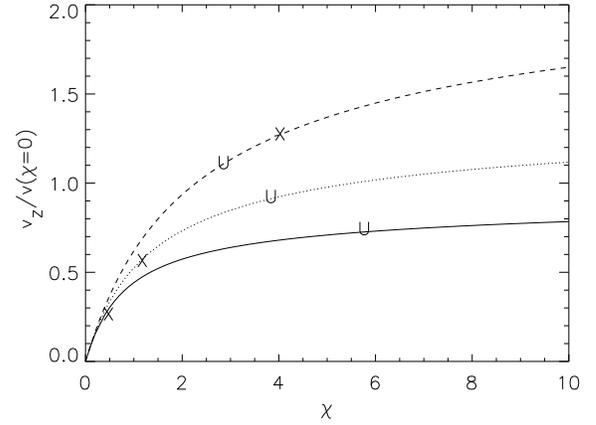}
\caption{\label{fig:f4} The ratio of $v_{\rm z}$ to the Kelperian velocity at the
footprint of the flow, plotted against $\chi$, for launch angles
$\theta_{\rm o} = 20^{\rm o}$ (solid), $30^{\rm o}$ (dotted), and $40^{\rm o}$
(dashed). The Alfv\'{e}n critical points (X) and position of the UFO (U) are
indicated for all three. Note that for $\theta_{\rm o}= 40^{\rm o}$, the UFO is
at a sub-Alfv\'{e}nic point.}
\end{figure}

\begin{figure}[htb]
\hspace{-1.0cm}\includegraphics[scale=0.35,angle=180]{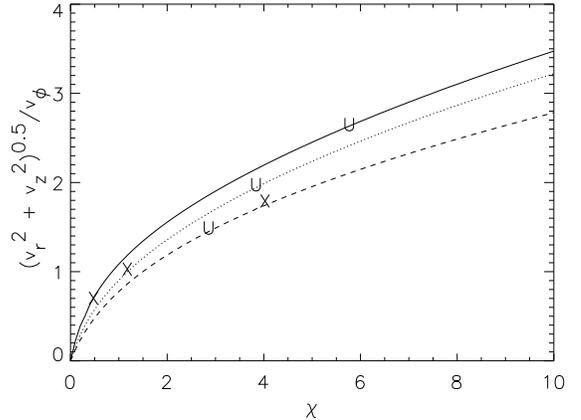}
\caption{\label{fig:f5} The ratio of the poloidal to the azimuthal components of velocity,
plotted against $\chi$, for all $\theta_{\rm o}$, as in Figure 4. The
Alfv\'{e}n critical and UFO positions are indicated.}
\end{figure}

\subsection{NGC 4151: a Case Study}

While MHD models have been successful in predicting the general properties of
warm absorbers \citep[e.g.,][]{proga2000} and UFOs \citep{fukumura2014}, there have not been
any tight constraints on critical parameters, such as density, location, and
magnetic field strengths, based on observational analysis. We are now able to do so
in the case of NGC 4151, for which we have constraints on the properties of 
the UFO and the inclination of the accretion disk.

Based on the constraint
$\Delta r$/$r < 1$, the maximum radial distance for the UFO in NGC~4151 is $r
= 3.2 \times 10^{15}$ cm (see Table 1).
We also use an inclination of $\sim$ 45$^{\rm o}$ \citep{das2005} and a black hole mass $M = 4.57 \times 10^{7} M_\odot$ \citep{bentz2006}.

\subsubsection{Predicted Properties of the Flow}

To characterize the flow, we use the method outlined in Section 3.1 to calculate the
foot-print radius as a function of launch angle and the
flow scaling parameters at the position of the UFO. 
In Table 3, we give the values of parameters characterizing the flows for
launch angles $\theta_{\rm o} =$ 20$^{\rm o}$, 30$^{\rm o}$, and 40$^{\rm o}$.
For a given inclination, as launch angle increases, the radial distance of the
footprint increases. However, the position along the flow where the streamline 
intersects our line-of-sight decreases, as evidenced by the decrease in
$\chi_{\rm UFO}$. In Figures 3 through 9, the physical parameters of the flow for the different values
of  $\theta_{\rm o}$, are plotted against $\chi$. 

As shown in Figure 3, the ratio of $v_{\rm z}$ 
to the Kelperian velocity at the footprint increases with $\chi$ along the streamline, eventually
reaching the asymptotic values, $f_\infty$, listed in Table 3.
In Figure 4, the ratio of the poloidal component of the flow, $\sqrt{v_{\rm r}^2+v_{\rm z}^2}$, to the
azimuthal component, $v_{\phi}$, as a function of
$\chi$, is shown for
$\theta_{\rm o} = 20^{\rm o}, 30^{\rm o}$ and $40^{\rm o}$. 
For each $\theta_{\rm o}$, the flow becomes increasingly poloidal with increasing
$\chi$. Note that for $\theta_{\rm o} =  40^{\rm o}$, the outflow is sub-Alfv\'{e}nic at the 
position of the UFO. This is the result of a greater magnetic field strength 
compared to the other cases (see below).

The density, $\rho$, relative to that at the Alfv\'{e}n critical point, is shown
in Figure 5. 
As shown in EBS92, $\rho$ decreases with increasing $m$. For a given $\chi$, $m$ is smaller for
larger $\theta_{\rm o}$. This is largely due to the value of $\kappa$ (see, Table 3 and EBS92, Equation 3.15), 
hence, $\rho$/$\rho_{\rm A}$ is greater for smaller launch angles. This is clearly evident for $\theta_{\rm o} = 40^{\rm o}$, when the
UFO lies below the Alfv\'{e}n critical point, as noted above. 
In Figure 6, we plot $P_{\rm mag}$ relative to that at the Alfv\'{e}n critical point. Again, 
the values for different $\theta_{\rm o}$ can be understood
in terms of $m$ and the proximity of the UFO relative to the critical point.

Since the flow velocity depends on the magnetic field strength, it is
instructive to compare the magnetic pressure with energy density. \citet{rees1987}
suggested that the broad emission-line region (BLR) clouds in AGN are magnetically confined
and, furthermore, that there is equipartition between the magnetic field and
gravity. If so, the magnetic pressure and the gravitational energy density
should be roughly equal, or
\begin{equation}
 P_{\rm mag}  \approx \frac{GM\rho}{r}.
\end{equation}
In its role in cloud-confinement, the $B$ field is not directly affecting
the BLR dynamics. However, in the case of an MHD wind, the $B$ field is the
mechanism that drives the outflow, and, therefore, there is no reason to expect
equipartition. 

In Figures 7 - 9,
we show the ratio of $P_{\rm mag}$ to gravitational energy density.
In each case, conditions are close to equipartition near the footprint. For
$\theta_{\rm o} = 20^{\rm o}$, the ratio rises, but the flow stays close to 
equipartition. However, at greater $\theta_{\rm o}$ the ratio rapidly exceeds 
equipartition. This is consistent with the corresponding greater $B$ field strengths (see
Table 4).

A larger $\theta_{\rm o}$ means that the
field lines are more coiled on top of one another. The field must be 
strong enough to buoy a vertical column of outflowing material in the MHD wind.
As shown in Table 4, all components of the $B$ field are greater at the
position of the UFO,
the toroidal component, $B_{\phi}$, increases the most with larger $\theta_{\rm o}$, as
expected.

In Figures 7-9, we also show the ratio of $P_{\rm mag}$ to the kinetic energy density, $
\frac{1}{2}\rho v_{\rm tot}^2$, where $v_{\rm tot}^2$ is the sum of the squares
of the poloidal and toroidal components of velocity. For each $\theta_{\rm o}$, the ratio
is relatively flat, which is consistent with the dependence of both quantities
on $m$.  Finally, in Table 4 we list the log of the ratio of $P_{\rm mag}$ to
the gas pressure, $P_{\rm gas}$, 
at the position if the UFO. In each case, $P_{\rm mag} >> P_{\rm gas}$, which is consistent with a
cold MHD flow. 
 
\begin{deluxetable*}{llllllll}
\scriptsize
\tablecaption{Model Predicted Properties}
\tablenum{1}
\tablewidth{8pt}
\tablehead{
\colhead{name}
&\colhead{r}
&\colhead{log Xi} &\colhead{logN$_{H}$}
&\colhead{logn$_{H}$}
&\colhead{log T} &\colhead{HC$^a$} &\colhead{FM} \\
\hline
\colhead{}
&\colhead{cm}
&\colhead{}
&\colhead{cm$^{-2}$}
&\colhead{cm$^{-3}$}
&\colhead{K}
&\colhead{\%}
&\colhead{}}
\colnumbers
\startdata
&&& UFOs &&&&\\
\hline
NGC 4151 & 15.5 & 4.33 & 22.9 & 7.4 & 7.32 & 96.5  & 1.01 \\
IC 4329a & 16.1 & 4.87 & 23.1 & 7.1 & 7.45 & 99.0 & 1.00\\
Mrk 509 obs1 & 15.8 & 5.16 & 23.9 & 7.5 & 7.49 & 99.4 & 1.00\\
Mrk 509 obs2 & 15.7 & 5.15 & 23.6 & 7.7 & 7.49 & 99.4 & 1.00\\
Mrk 509 obs3 & 16.5 & 4.25 & 23.7 & 7.3 & 7.28 & 95.1 & 1.01\\
Ark 120 & 16.3 & 4.55 & 23.7 & 7.5  & 7.39 & 97.8 & 1.00\\
Mrk 79 & 16.5 & 4.19 & 23.1 & 6.7  & 7.25 & 94.8 & 1.01\\
NGC 4051 obs1 & 14.9 & 4.37 & 23.0 & 8.1  & 7.34 & 96.9 & 1.01\\
Mrk 766 obs1 & 17.1 & 3.73 & 22.4 & 5.3  & 6.88 & 81.3 & 1.06\\
Mrk 766 obs2 & 15.9 & 4.28 & 23.0 & 7.3  & 7.29 & 96.0 & 1.01\\
Mrk 841 & 17.0 & 3.91 & 23.0 & 6.0 & 7.05 & 88.5  & 1.03\\
Mrk 290 & 16.3 & 3.91 & 23.4 & 7.1  & 7.05 & 87.9  & 1.03\\
Mrk 205 & 15.7 & 4.62 & 23.8 & 8.1  & 7.42 & 98.1 & 1.00\\
MCG$-$5$-$23$-$16 & 16.6 & 4.25 & 22.7 & 6.1  & 7.28 & 95.7 & 1.01\\ 
NGC 4507 & 15.9 & 4.53 & 23.0 & 7.1 & 7.39 & 97.9 & 1.00\\
\hline
&&& non-UFOs &&&&\\
\hline
Mrk 279 & 17.9 & 3.33 & 22.2 & 5.0 & 6.34 & 46.9 & 1.32\\
NGC 3516 obs1 & 16.6 & 3.73 & 22.5 & 5.9  & 6.87 & 81.3 & 1.06\\
NGC 3516 obs2 & 16.8 & 3.76 & 22.5 & 6.4 & 6.90 & 82.7 & 1.05\\ 
NGC 3783 obs1 & 18.0 & 3.13 & 22.5 & 4.5 & 6.09 & 24.4 & 1.73\\
NGC 3783 obs2 & 17.8 & 3.23 & 22.4 & 4.6 & 6.20 & 33.7 & 1.45\\
NGC 3783 obs3 & 18.0 & 3.13 & 22.5 & 4.5 & 6.09 & 24.4 & 1.73\\
ESO 323-G77 & 17.0 & 3.53 & 23.5 & 6.5 & 6.62 & 65.3  & 1.12 \\
\hline
\enddata
\tablenotetext{a}{HC is the fractional contribution of Compton Heating.}
\end{deluxetable*}
\normalsize

\begin{deluxetable*}{llll}
\scriptsize
\tablecaption{Log Fe Column Densities}
\tablenum{2}
\tablewidth{8pt}
\tablehead{
\colhead{name}
&\colhead{Fe$^{+23}$}
&\colhead{Fe$^{+24}$(mes$^a$)} 
&\colhead{Fe$^{+25}$(mes$^a$)}\\
\hline
\colhead{}
&\colhead{cm$^{-2}$}
&\colhead{cm$^{-2}$}
&\colhead{cm$^{-2}$}}
\startdata
&  UFOs & & \\ 
\hline
NGC 4151 & 14.4 & 16.6 & 17.7~(17.9)\\
IC 4329a & 12.8 & 15.6  & 17.4~(17.3) \\
Mrk 509 obs1 & 13.0 & 16.0 & 18.0~(17.9)\\
Mrk 509 obs2 & 12.5  & 15.5  & 17.6~(17.6) \\
Mrk 509 obs3 & 15.7 & 17.7 & 18.7  \\
Ark 120 & 14.6 & 17.0  & 18.4  \\
Mrk 79 & 15.2 & 17.2 & 18.1~(18.0) \\
NGC 4051 obs1 & 14.3 & 16.6  & 17.8 \\
Mrk 766 obs1 & 16.1 & 17.4  & 17.7~(18.0) \\
Mrk 766 obs2 & 14.6 & 16.7 & 17.8~(17.8) \\
Mrk 841 & 16.1 & 17.7  & 18.2 \\
Mrk 290 & 16.6 & 18.1 & 18.6 \\
Mrk 205 & 14.5 & 17.0  & 18.4~(18.3) \\
MCG$-$5$-$23$-$16 & 14.5 & 16.6  & 17.6~(17.6)\\ 
NGC 4507 & 13.8  & 16.2  & 17.6  \\
\hline
&  non-UFOs & & \\ 
\hline
Mrk 279 & 17.1 & 17.5~(17.5) & 17.0 \\
NGC 3516 obs1 & 16.2 & 17.5~(17.5) & 17.8~(17.8)\\
NGC 3516 obs2 & 16.1 & 17.5~(17.3) & 17.8~(17.7)\\
NGC 3783 obs1 & 17.3 & 17.4~(17.4) & 16.5  \\
NGC 3783 obs2 & 17.3 & 17.6~(17.6) & 16.9  \\
NGC 3783 obs3 & 17.3 & 17.4~(17.4) & 16.5 \\
ESO 323-G77 & 18.0 & 18.8~(18.8) & 18.6~(18.4)\\
\hline
\enddata
\tablenotetext{a}{mes refers to the column densities derived from
curve-of-growth (Section 2.1).}
\end{deluxetable*}
\normalsize

\begin{deluxetable*}{lcccccccc}
\scriptsize
\tablecaption{Flow Parameters for NGC 4151}
\tablenum{3}
\tablewidth{8pt}
\tablehead{
\colhead{$\theta_{\rm o}$}
& \colhead{$\chi_{\rm UFO}$}
& \colhead{$\chi_{\rm A}$} 
& \colhead{$f_{\rm UFO}$}
& \colhead{$f_{\infty}$}
& \colhead{$\lambda$} 
& \colhead{$\kappa$}
& \colhead{log($r_{\rm o}$)} 
& \colhead{log($v_{\rm o}$})\\
\colhead{}
& \colhead{}
& \colhead{} 
& \colhead{}
& \colhead{}
& \colhead{} 
& \colhead{}
& \colhead{(cm)} 
& \colhead{(cm s$^{-1}$})} 
\colnumbers
\startdata
20$^{\rm o}$ & 5.64 & 0.36 & 0.73 & 0.98 & 2.95 & 2.10 & 14.59 & 9.59\\
30$^{\rm o}$ & 3.71 & 1.06 & 0.91 & 1.46 & 4.68 & 0.65 & 14.77 & 9.51\\
40$^{\rm o}$ & 2.73 & 3.91 & 1.08 & 2.42 & 10.3 & 0.14 & 14.91 & 9.43\\
\enddata
\end{deluxetable*}
\normalsize

\begin{deluxetable*}{lccccc}
\footnotesize
\tablecaption{Magnetic Properties$^{a}$}
\tablenum{4}
\tablewidth{10pt}
\tablehead{
\colhead{$\theta_{\rm o}$}
& \colhead{$B_{\rm z}^b$}
& \colhead{$B_{\rm r}$} 
& \colhead{$B_{\phi}$}
& \colhead{$B_{\rm tot}$}
& \colhead{log(P$_{mag}$/P$_{\rm gas}$})}
\colnumbers
\startdata
20$^{\rm o}$ & 12.9 & 6.2 & 46.3 & 48.4 & 2.76\\
30$^{\rm o}$ & 31.6 & 14.7 & 110.3 & 115.7 & 3.52 \\
40$^{\rm o}$ & 83.6 & 36.5 & 360.4 & 371.8 & 4.53 \\
\enddata
\tablenotetext{a}{Evaluated at r$_{\rm UFO}$.}
\tablenotetext{b}{$B_{\rm i}$ is the i-component of the magnetic field in units of G.}
\end{deluxetable*}
\normalsize

\begin{figure}
\hspace{-1.0cm}\includegraphics[scale=0.35,angle=180]{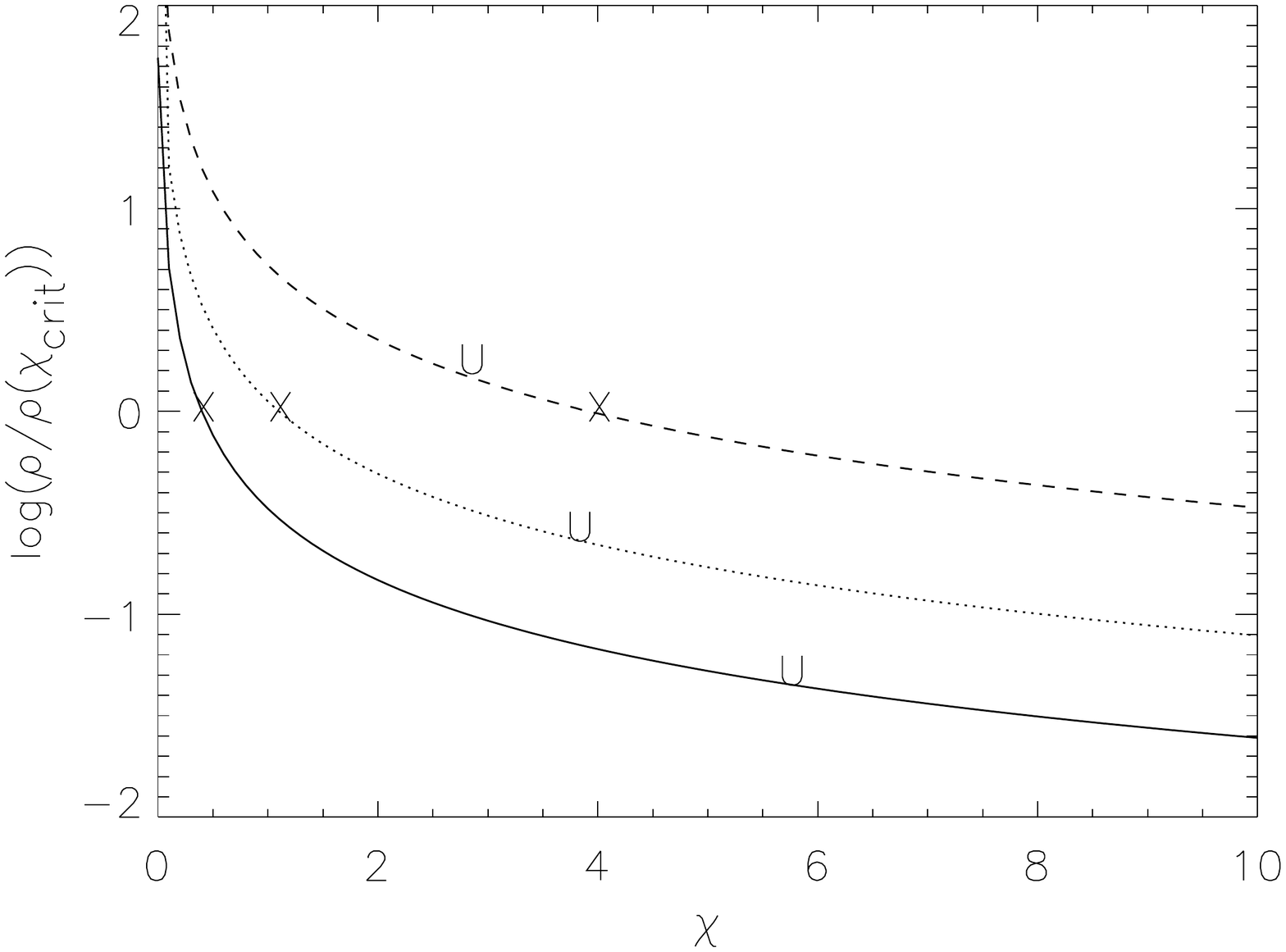}
\caption{\label{fig:f8} Density relative to that at the Alfv\'{e}n critical point,
as a function of $\chi$ for 
$\theta_{\rm o} = 20^{\rm o}$ (solid), $30^{\rm o}$ (dotted), and $40^{\rm o}$
(dashed).}
\end{figure}

\begin{figure}
\hspace{-1.0cm}\includegraphics[scale=0.35,angle=180]{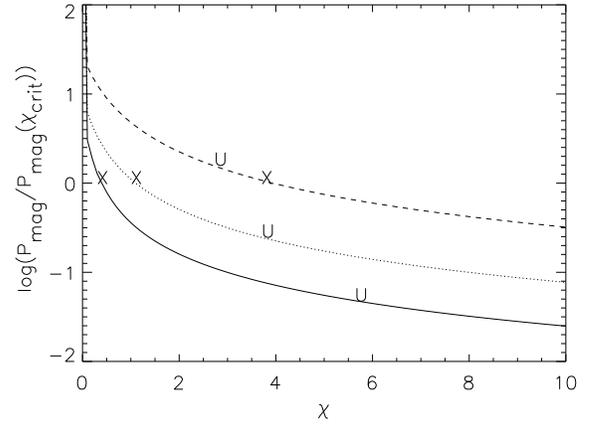}
\caption{\label{fig:f9} Magnetic pressure relative to that at the Alfv\'{e}n
critical point, as a function of $\chi$, for $\theta_{\rm o} = 20^{\rm o}$ (solid), $30^{\rm o}$ (dotted), and $40^{\rm o}$
(dashed). }
\end{figure}

\begin{figure}[htb!]
\hspace{-1.0cm}\includegraphics[scale=0.35,angle=180]{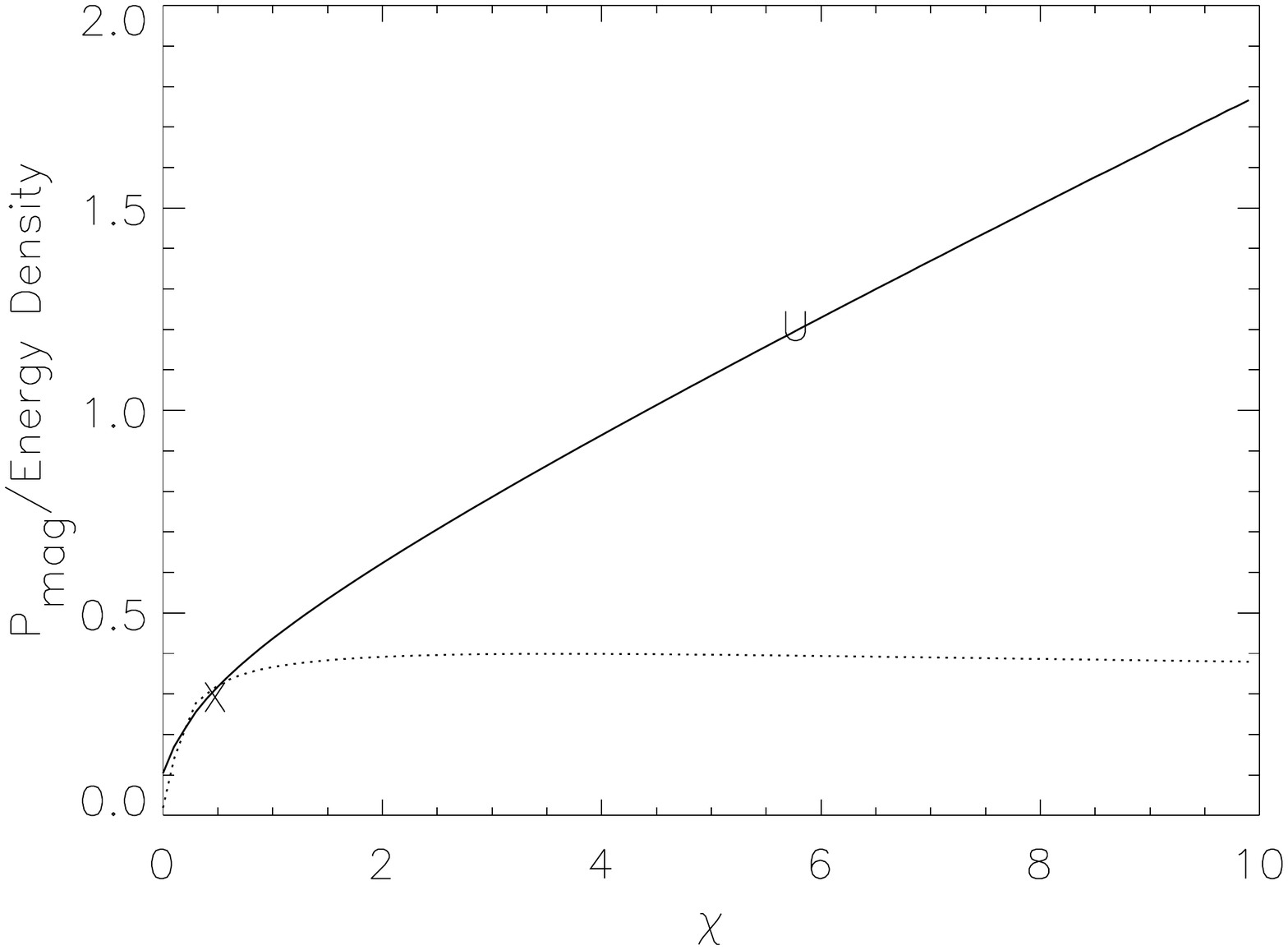}
\caption{\label{fig:f10} The ratios of magnetic pressure to gravitational energy
density (solid) and kinetic energy density (dotted), as a function of $\chi$ for
$\theta_{\rm o} = 20^{\rm o}$.}
\end{figure}

\begin{figure}[htb!]
\hspace{-1.0cm}\includegraphics[scale=0.35,angle=180]{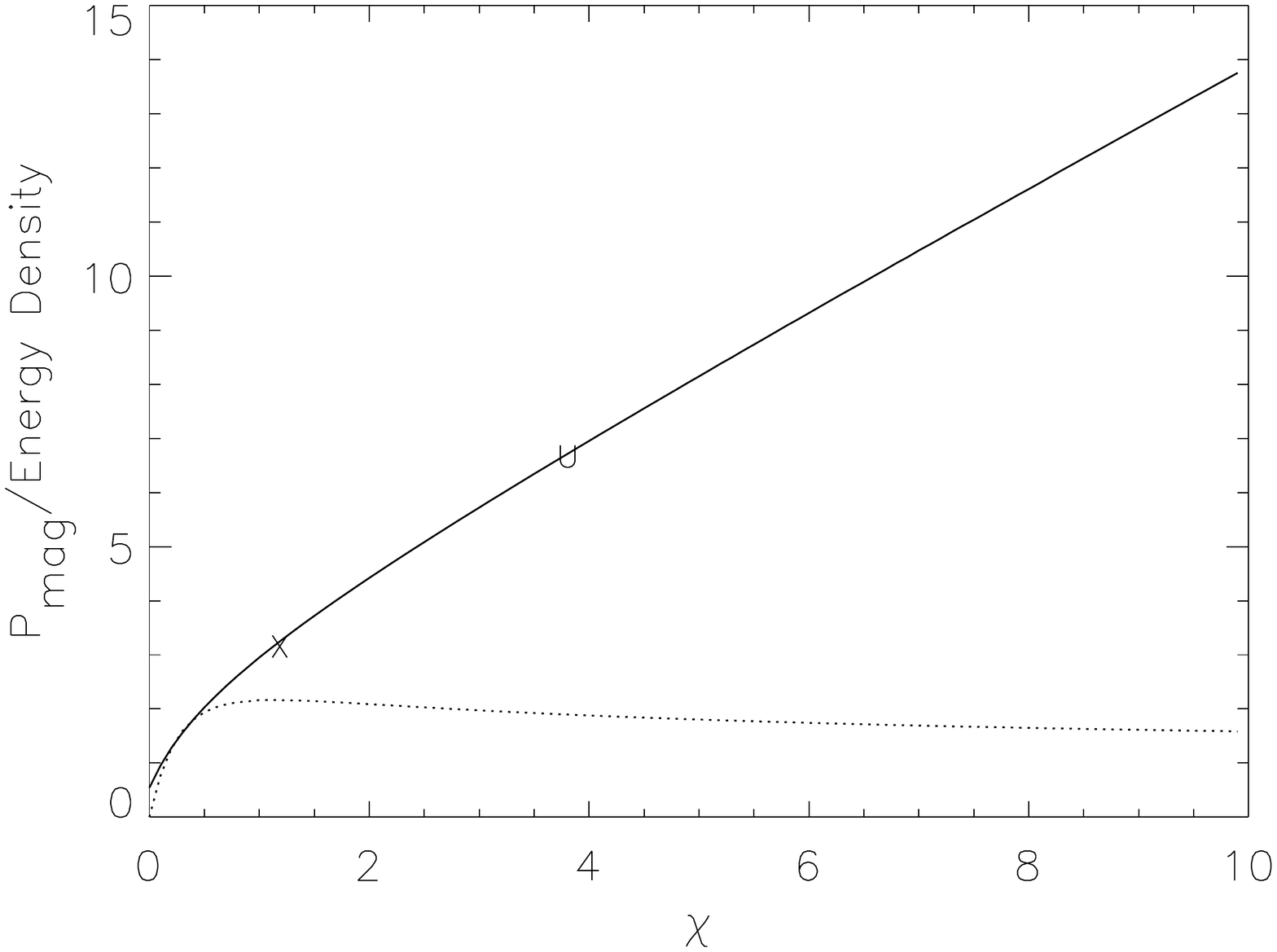}
\caption{\label{fig:f11} The ratios of magnetic pressure to gravitational energy
density (solid) and kinetic energy density (dotted), as a function of $\chi$ for
$\theta_{\rm o} = 30^{\rm o}$.}
\end{figure}

\begin{figure}[htb!]
\hspace{-1.0cm}\includegraphics[scale=0.35,angle=180]{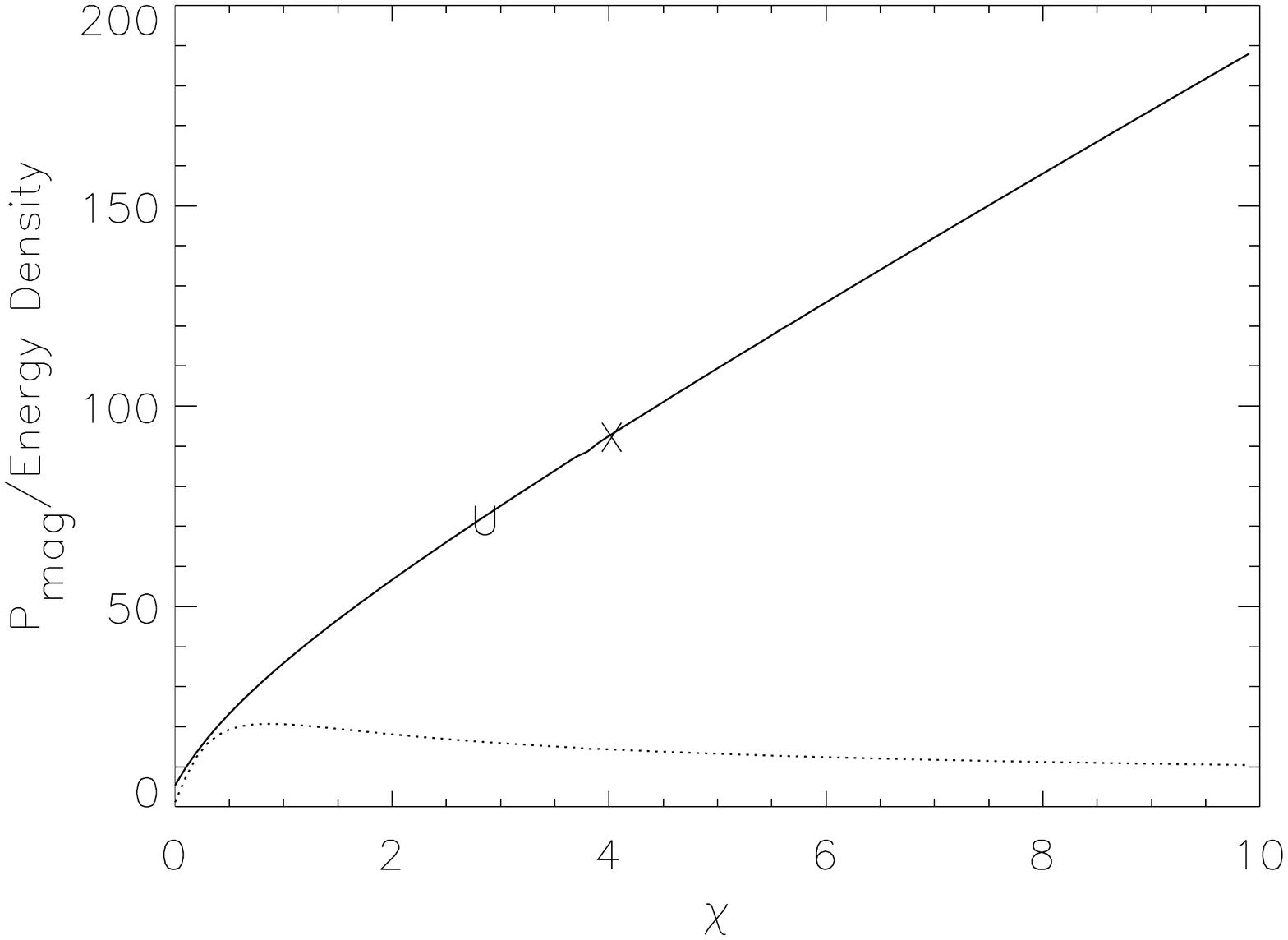}
\caption{\label{fig:f12} The ratios of magnetic pressure to gravitational energy
density (solid) and kinetic energy density (dotted), as a function of $\chi$ for
$\theta_{\rm o} = 40^{\rm o}$.}
\end{figure}

\subsubsection{Constraints on Velocity Structure}

 The UFO detected in the {\it XMM} observation of NGC~4151 was resolved, with a dispersion
 of $\sigma = 5.1 (+1.8/-1.4) \times 10^{3}$ km s$^{-1}$, or
 Full~Width at Half~Maximum ($FWHM$) $= 1.2 (+0.4/-0.3) \times 10^{4}$ km s$^{-1}$ \citep{tombesi2010}. 
 As noted above, \citet{tombesi2011} fit the
 spectrum with XSTAR models by including micro-turbulence. \citet{bottorff&ferland2000} suggested that the smoothness 
 of BLR emission-lines were the result
 of micro-turbulence and that turbulent BLR clouds could exist if magnetically confined. 
 Hence, in the presence of strong $B$ fields, such as those calculated in the MHD modeling,
 it is possible that the UFOs are highly turbulent.
 
On the other hand, a large $FWHM$ can result from a large velocity gradient along our line-of-sight through the 
absorbing material. This scenario was discussed in detail in
\citet{bottorff2000}. Here we use their Equation 14,
reformatted in terms of $\Delta r$/$r$. This allows for more direct comparison
of the MHD
model kinematics to our photo-ionizing model results. The relationship is:

 \begin{equation}
   \frac{\Delta r}{r}~=~4\Big{(}\frac{\Delta v}{2v_{r}}\Big{)}\Big{[}1~-~\Big{(}\frac{\Delta v}{2v_{r}}\Big{)}^{2}\Big{]}^{-2} 
 \end{equation}
where $\Delta v$ is $FWHM$, and $v_{r}$ is the radial velocity of the UFO. From
this, $\Delta r$/$r$ $\approx$ 0.8, which is consistent with the constraint on the
Cloudy models that $\Delta r$/$r$ $\leq$ 1 (see Section 2.1). This can occur if
our line-of-sight passes through streamlines originating at different launch radii \citep[see][Figure 1]{bottorff2000}. We suggest that the
observed $FWHM$ is more likely due to a radial velocity gradient, rather than micro-turbulence. 

\section{Discussion}

As shown in the previous section, the UFO in NGC 4151 can be characterized as part
of a cold MHD flow, with an origin in the accretion disk. Mass outflow
has been well-studied in NGC~4151  \citep[e.g.,][]{crenshaw2015}, and we have constraints
on the physical conditions and radial distances of the various components of
absorption \citep{kraemer2005}. Therefore, we are able to consider the UFO in the context of mass
outflow in this source.

As discussed in \citet[]{kraemer2005, kraemer2006} and \citet{couto2016}, there are
two main components of absorption in NGC 4151: ``XHIGH'', which was initially
detected by the presence of Mg~XII, S~XIV, and S~XVI absorption lines, and
D$+$Ea, which causes the broad-band soft X-ray absorption and has a UV signature
in the form of saturated C~IV, N~V, and O~VI lines. Even though most of the {\it
Chandra} and XMM-{\it Newton} observations of NGC~4151 found the source in
very low-flux states, with L$_{\rm bol}$/L$_{Edd}$ $\sim$ a few percent, 
\citet{couto2016} demonstrated that D$+$Ea could be radiatively accelerated. 
However, XHIGH was too highly ionized for radiative acceleration in a
sub-Eddington source, hence it could be MHD-driven. 

\citet[][and references therein]{fukumura2014} suggest that an MHD-driven disk wind will
result in a continuous distribution of $N_{\rm H}$ per decade of ionization parameter,
which results in a density law of the form $n(r) \propto r^{-\alpha}$, where
$\alpha \sim 1$. 
As discussed in \citet{couto2016}, the similar values of $N_{\rm H}$ for D$+$Ea, XHIGH, and
the UFO are consistent with this scenario. However, the density and location of D$+$Ea relative
to the UFO yields $\alpha \sim 0.5$, which is inconsistent with \citet{fukumura2014}'s requirement
for MHD. Based on photo-ionization modeling, \citet{kraemer2005} argued that XHIGH
must be closer to the continuum source than D$+$Ea, and \citet{couto2016} determined that the conditions in XHIGH 
are in agreement with the MHD model proposed by \citet{fukumura2014} Overall, this picture is suggestive
of stratification of the outflow in which the interior sections are MHD-driven while, at sufficiently
large radial distances, radiation-driving dominates. Also, XHIGH's properties overlap the lower ionization
end of the non-UFOs. Hence, one can envision a scenario in which UFOs are launched at the smallest radii and non-UFOs,
while still MHD-driven, form further out. This is consistent with our photo-ionization modeling analysis (see Table 1).

Comparing XMM-{\it Newton} \citep[]{tombesi2010, tombesi2011} and
{\it Suzaku} \citep{gofford2015} observational results, there are cases of large
velocity differences occurring on relatively short timescales, For example, for the UFO in
NGC~4151, \citet{tombesi2010} found $v_{\rm obs}/c = 0.106 \pm 0.007$, while, in a {\it
Suzaku} spectra taken $\sim 18$ days later,  \citet{gofford2015}
found $v_{\rm obs}/c = 0.055 \pm 0.023$. A more extreme difference was seen
in Mrk 279 , with $v_{obs}/c < 0.007$ \citep{tombesi2011} versus $0.222 \pm 0.006$,
approximately 3.5 yrs later \citep{gofford2015}.  In the case of NGC~4151, the
difference in $v_{\rm obs}$ might be consistent with a change in the direction
of the velocity vector, as suggested for a component of UV absorption
in NGC 3783 \citep{gabel2003}, but such a scenario would require a small covering
factor for the UFO, which seems unlikely given its large column density and the
possibility that our line-of-sight passes through different streamlines, as
discussed above. We suggest that it is more plausible that these are individual
components of absorption, whose velocity differences result from different 
launch radii or different physical conditions, such as the magnetic field strengths, at
the times of ejection.   

Other than NGC~4151, there do not appear to be sources that harbor UFOs and non-UFOs, at least
in the same epoch. Based on our model constraints, the non-UFOs are at larger distances (see Table 1),
which implies that the may have originated at larger $r_{\rm o}$. In the context of an MHD outflow, the 
lower values of $v_{\rm r}$ are consistent with lower Kelperian velocities at the launch points, hence
lower outflow velocities. However,
this does not explain why the two types cannot be present in the same objects. One possiblity is
that the conditions in the disk are such that either UFOs or non-UFOs are created. If this
is related to magnetic field strength, there may be associated changes in the core radio emission.

\citet{crenshaw2012} found that $L_{\rm KE} = 0.25~{\rm to}~1.6~\times 10^{41}$ erg s$^{-1}$, or 
$0.34~{\rm to}~2.0~\times 10^{-3} L_{\rm bol}$, for the combined UV and X-ray absorbers in NGC~4151.
Including the optical/UV NLR emission line gas increases
$L_{\rm KE}$ to a peak of 4.3 $\times 10^{41}$ erg s$^{-1}$ (0.006 - 0.008 of $L_{\rm
bol}$). This is barely sufficient for feedback. Based on our characterization
of the UFO, we obtain $L_{\rm KE} = 3.5 \times 10^{43}$ erg s$^{-1}$, for
$C_{\rm g} = 0.5$, which
is on the same order as $L_{\rm bol}$. Therefore, if the UFO has a large
covering factor, and can maintain its integrity as it moves into the galactic
bulge, it has sufficient kinetic luminosity for effective AGN
feedback.
 
If MHD-driven UFOs play an important role in AGN feedback, the interaction between the SMBH and the host
galaxy is via the magnetic properties of the disk. This is opposed to feedback due to radiatively-driven
winds, which is the more typically invoked scenario. Interestingly, the cold MHD model discussed by BP82 was
intended to explain radio jets. Therefore, the form of ``UFO feedback'' we describe in the present paper is simply
a less energetic form of the same phenomenon. Since UFOs can form in sub-Eddington sources, which do not seem
to be able to produce the high $L_{\rm KE}$ winds required for feedback, perhaps a more
broadly defined radio mode feedback, which includes UFOs, is the dominant means of SMBH/host interaction.    

\section{Conclusions}

Starting with a sample of AGN with intrinsic Fe~XXV and Fe~XXVI absorption 
detected by XMM-{\it Newton} \citep[]{tombesi2010, tombesi2011}, we have analyzed the physical 
conditions within the absorbers, using photo-ionization models generated with
Cloudy \citep{ferland2013}. We have determined the following:

1. It has been shown that there is a continuum of properties, with
decreasing ionization and (generally) column density, over the range from
UFOs to non-UFOs to warm absorbers \citep[e.g.,][]{tombesi2011}.  The highest ionization 
warm absorbers appear to overlap the non-UFO region, 
which suggests there may be some connection between these phenomena. We have shown that UFOs and non-UFOs
occupy different regions on an S-curve, with little overlap, with the former in
the Compton-dominated range, while the latter are in the vertical range, where other cooling
mechanisms become important. 
Based on our model constraints, the non-UFOs lie at greater radial distances than the UFOs. Overall, this
is consistent with $n_{\rm H} \propto r^{-\alpha}$, with
$\alpha < 2$, in which case the ionization state of the absorbers decreases with distance,
with an associated change in the heating and cooling processes.

2. Cloudy models predict that UFOs and non-UFOs are characterized by FM $\sim$ unity
hence they are too highly ionized to be radiatively accelerated in sub-Eddington sources,
unless the UFOs were in a much lower ionization state at their launch points. This
suggests another means of acceleration, such as an MHD-driven flow. 

3. To explore the possibilty of MHD-driving, we applied the cold MHD model 
detailed in BP82 and EBS92 to case of NGC~4151, for which the inclination of the
black hole/accretion disk has been constrained \citep{das2005}. Specifically,
we followed the flow parameterization in EBS92, for which the poloidal part of the streamlines are 
parabolic, with a footprint in the accretion disk.  For a range of
launch angles, we find that the observed velocity is
consistent with MHD acceleration along a streamline and we are able to trace 
the origin of the UFO back to its footprint radius.
Also, with this geometry, the observed the relationship between the $FWHM$ and 
$v_{\rm obs}$ of the UFO in NGC~4151 is consistent with a velocity gradient 
through different flow streamlines. 

4. For NGC~4151 we have been able to constrain the magnetic field strength and magnetic
pressure in the flow. We find that magnetic pressure far exceeds gas
pressure predicted by the Cloudy models, consistent with the
definition of a cold flow. Also, the magnetic pressure generally exceeds the gravitational
energy density, therefore equiparition does not apply. 

Given the simplicity of this model, such as the assumptions of rigid field lines and
parabolic geometry, we do not suggest that these results be taken as the 
final word on MHD-driven outflows. Rather, the physical parameters, such as density
and magnetic
field strength along the streamlines, will be useful inputs to more sophisticated
models, such as those developed by \citet{fukumura2014}. However, if, as we
suggest, these outflows are magnetically driven disk-winds, understanding
their variabilty and the different origins of UFOs and non-UFOs may provide
new insight into the physics of accretion disks in AGN. It would be particularly interesting
if there was a connection between UFO properties and the radio emission from these
objects. Finally, if UFOs are an important feeback mechanism, particularly in
sub-Eddington AGN, it implies that magnetic properties of the disk can affect the host galaxy,
as is the case for radio-mode feedback.

\section{Acknowledgments}

We thank E. Behar, D. Kazanas, K. Fukumura, C. Shrader, D. Proga, and M. Rees
for useful comments and suggestions. We also thank G. Ferland and associates
for their continuing maintenance of Cloudy. F.T. acknowledges support from the
Programma per Giovani Ricercatori - anno 2014 ``Rita Levi Montalcini''.

\bibliographystyle{apj}      
\bibliography{ms_rev1.bbl}
   



\end{document}